\documentclass[smallextended]{svjour2}                    
\smartqed  
\usepackage{graphicx}
\usepackage{subfigure}
 \journalname{Journal of Statistical Physics}
\begin{document}

\title{The combined effect of connectivity and dependency links on percolation of networks
}
\subtitle{In honor of Professor Cyril Domb, our teacher}


\author{Amir Bashan         \and
        Shlomo Havlin 
}


\institute{A. Bashan \at
              Bar Ilan University, Ramat Gan, Israel \\
              \email{amir.bashan@gmail.com}           
           \and
           S. Havlin \at
              Bar Ilan University, Ramat Gan, Israel \\
              \email{havlin@ophir.ph.biu.ac.il}
}

\date{Received: date / Accepted: date}

\maketitle

\begin{abstract}
Percolation theory is extensively studied in statistical physics and mathematics with applications
in diverse fields.
However, the research is focused on systems with only one type of links, connectivity links.
We review a recently developed mathematical framework for analyzing percolation properties of
 realistic scenarios of
networks having links of two types, connectivity and dependency links.
This formalism was applied to study Erd$\ddot{o}$s-R$\acute{e}$nyi (ER)
 networks that include also dependency links.
 For an ER network with average degree $k$ that is composed of dependency clusters of size $s$,
 the fraction of nodes that belong to the giant component,
 $P_\infty$, is given by $ P_\infty=p^{s-1}\left[1-\exp{(-kpP_\infty)}  \right]^s $
where $1-p$ is the initial fraction of randomly removed nodes.
Here, we apply the formalism to the study of random-regular (RR) networks and find a formula for
the size of the giant component in the percolation process:
$P_\infty=p^{s-1}(1-r^k)^s$ where $r$ is the solution of $r=p^s(r^{k-1}-1)(1-r^k)+1$.
These general results coincide, for $s=1$, with the known equations for percolation in ER and RR networks respectively
without dependency links. In contrast to $s=1$, where the percolation transition is second order, for  $s>1$ it is of first order.
 Comparing the percolation behavior of ER and RR networks we find a remarkable
difference regarding their resilience.
We show, analytically and numerically, that in ER networks with low connectivity degree
 or large dependency clusters, removal of
even a finite number (zero fraction) of the network nodes will trigger
 a cascade of failures that fragments the whole network.
Specifically, for any given $s$ there exists a critical degree value, $k_{min}$, such that an ER network
with $k\leq k_{min}$ is unstable and collapse when removing even a single node. This result is in contrast to RR networks where
 such cascades and full fragmentation can be triggered only by removal of a finite fraction of nodes in the network.

\end{abstract}

\section{Introduction}
\label{intro}
Percolation is a primary model for disordered systems which is extensively studied
both in mathematics \cite{Essam,Kesten,Aizenman} and in statistical physics \cite{Kirkaptrick,BundeHavlinBook,Stauffer,Domb}
and have applications in diverse phenomena in different disciplines ranging
from porous materials \cite{Zallen} and branched polymers \cite{BundeHavlinPolimers} to forest
fires \cite{Fire1} and epidemics spreading \cite{Epidemics1}.
In network science
\cite{WattsNature,barasci,bararev,PastorXX,mendes,NewmanSIAM,caldarelli1,Reuven_book,caldarelli2,Vespignani,NewmanBook,Albert,Newman2006,Barthelemy}, percolation theory is particular useful in studying
robustness of networks to a failure of their components \cite{cohena,cohena2,gallosPRL,callawayPRL}.
For a given network, when nodes are removed with probability $1-p$, or occupied with probability $p$,
 the percolation theory
explains the existence of a critical probability, $p_c$, such
that for $p$ below $p_c$ the network is composed of isolated small
clusters but for $p$ above $p_c$ a giant cluster, of fraction $P_\infty$ of the network, spans the entire network.
The percolation process represents a phase transition of second-order where
$P_\infty$  (the order parameter) approaches continuously to zero when $p$ approaches $p_c$ from above.
The value of $p_c$ is considered as a measure for the network robustness.
When $p_c$ is smaller, the network is more robust since more nodes have to be removed from the network in order to
fragment it completely.
The results of percolation theory were useful, for example, to evaluate the resilience
of the internet against random \cite{cohena} or malicious attacks
\cite{cohena2,gallosPRL,callawayPRL,Holme,Schneider}
 or to develop efficient immunization strategies
\cite{Schneider,Pastor2002,Cohen2003,Chen}.

 Percolation theory assumes that the network elements are connected by
 one type of links, connectivity links. However, in real systems, usually
 there exist two types of links, connectivity links for the function of the network
 and dependency links which represent that the function of given elements
 depend crucially on others. Network models containing both connectivity and dependency links
 were introduced recently for two interdependent networks \cite{Buldyrev,Parshani_PRL}
 and for single networks \cite{Parshani_PNAS,Bashan}. For further recent studies of
 interdependent networks see Refs. \cite{Shao2011,GaoNON,Huang,Yanqing}.

 Here, we review the general formalism for analyzing a single network composed of connectivity
 and dependency links and the results obtained for Erd$\ddot{o}$s-R$\acute{e}$nyi (ER) type networks.
 We apply here the formalism to analyze random-regular (RR) networks composed of these two types of links.
 We also compare between the results of ER and RR networks and show that in the presence of
 dependency links, RR networks are significantly more robust compared to ER networks. This is in contrast to the case of no dependency links where ER are more robust than RR networks.

\begin{figure}[ht]
\centering
\subfigure[]{ 
\includegraphics[scale=0.15]{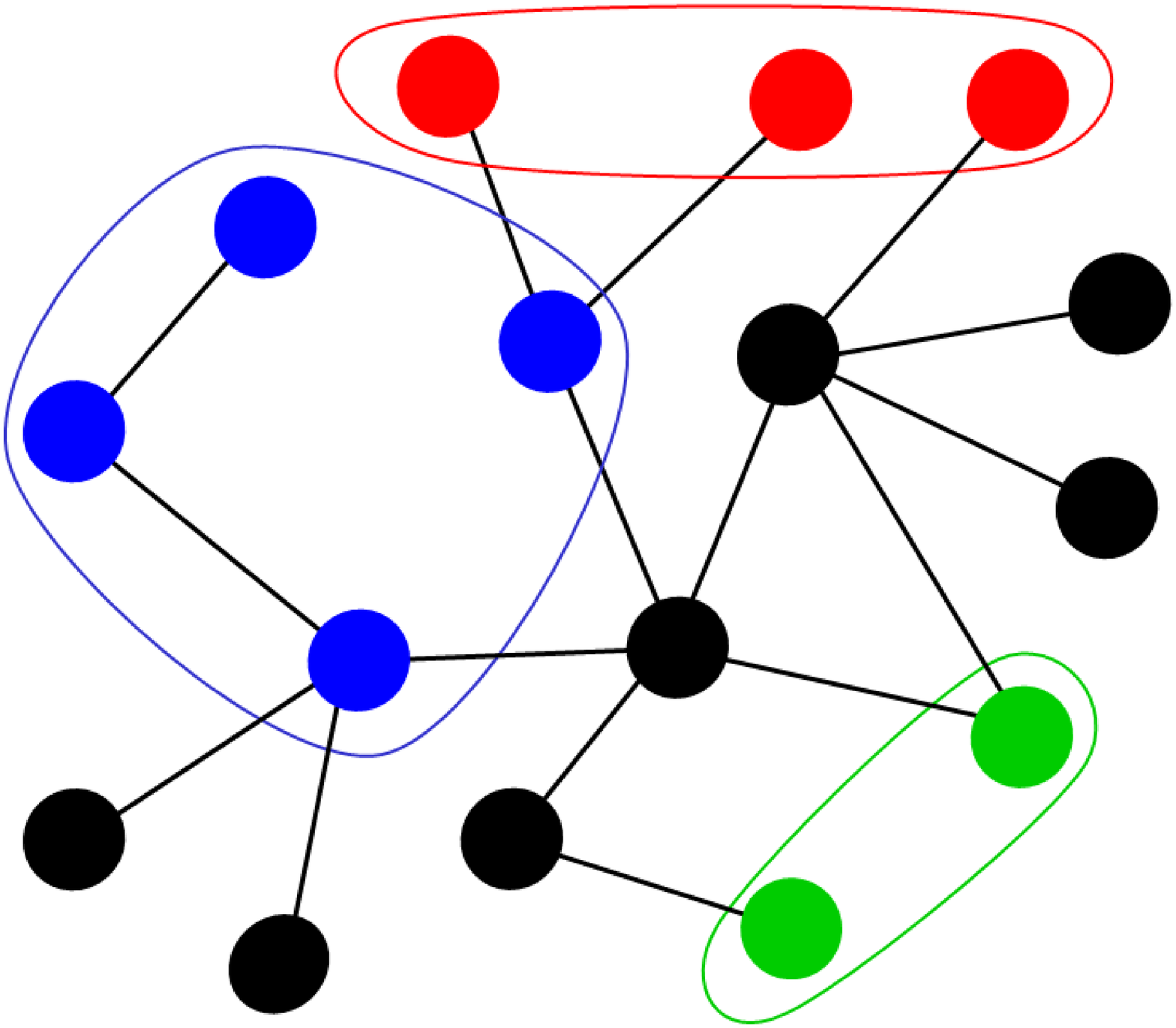}
\label{figure1_ER}
}
\subfigure[]{
\includegraphics[scale=0.15]{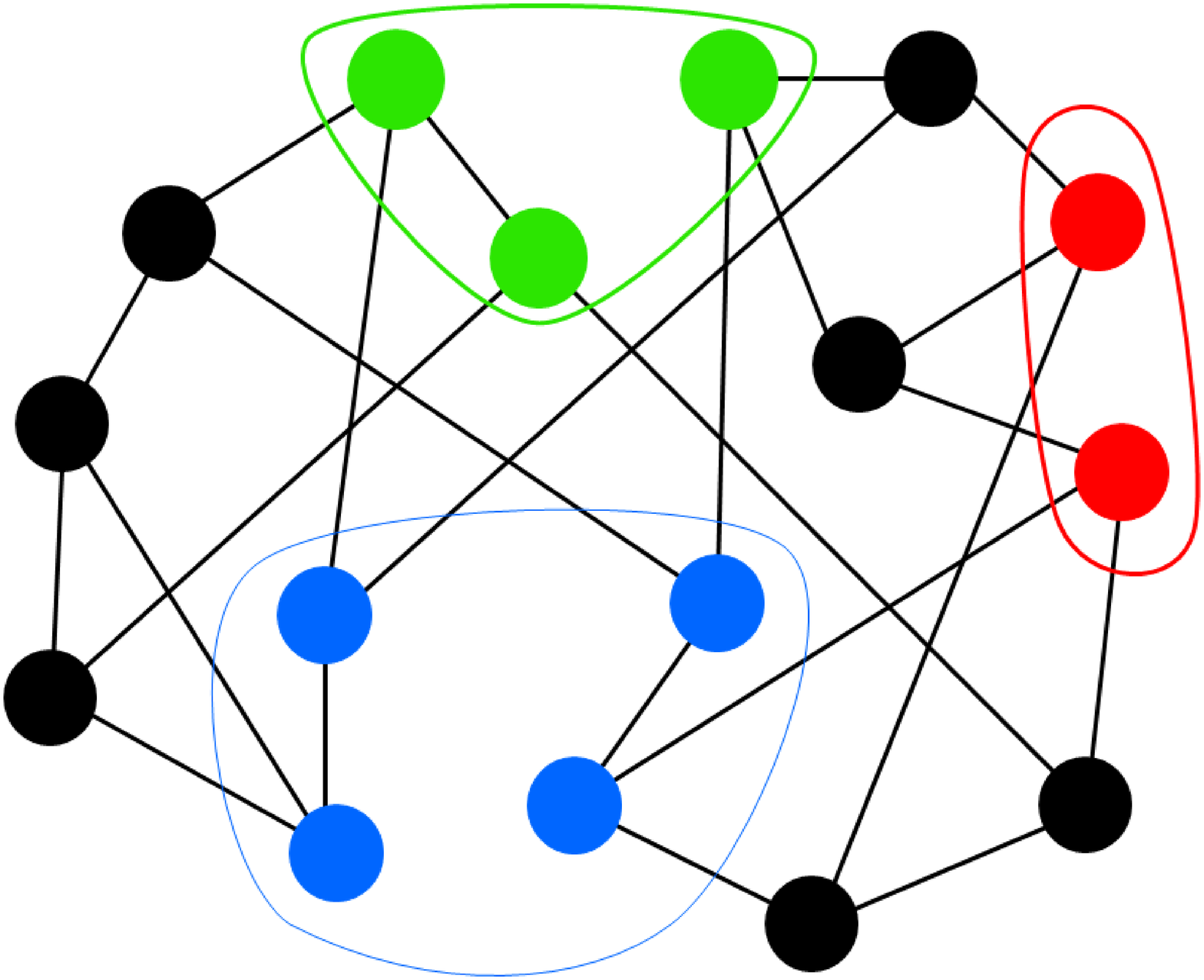}
\label{figure1_RR}
}
\caption{(Color online) Connectivity networks with dependency clusters. The edges represent
connectivity relations, while the (blue, red and green) groups surrounded by curves represent dependency
relations between all the nodes of the same group (color). The dependency relations
can be between very $``far"$ nodes in the connectivity network. (a) ER network with dependency clusters.
(b) RR network with dependency clusters where all nodes have same degree $k=3$.}
\label{figure1}
\end{figure}

\section{General formalism}
\label{sec:GenForm}
When nodes fail in a network containing both connectivity links and dependency clusters, two different processes occur.
(i) Connectivity links connected to the failed nodes also fail, causing other nodes to disconnect from the network (connectivity step).
(ii) A failing node cause the failure of all the other nodes of its dependency cluster,
 even though they are still connected via connectivity links (dependency step).
Thus, a node that fails in the connectivity step leads to the failure of its entire dependency cluster, which in turn leads to a new
connectivity step, which further leads to a dependency step and so on. Once the cascade process is triggered it will stop only if nodes that fail in one step do not cause additional failure in the next step.

In order to describe the network size during the cascade and, in particular, to find its size at steady state,
two functions are defined, $g_D(T)$ and $g_p(T)$, evaluating respectively the effect of the dependency clusters and the effect of
 the connectivity process on the network at each step of the cascade.
  After random removal of $1-T$ of the nodes, the size of the giant component is given by
 $g_p(T)$ and, equivalently, the part of the network that is not dependent on the removed nodes (and, thus, remain functional) is given
 by $g_D(T)$. Thus, when applying the dependency process on a network of size $x$
 the remaining functional nodes consisting  of a fraction $g_D(x)$, which is a fraction
 $\phi=x g_D(x)$  of the $N$ nodes in the original network.
Similarly, applying the connectivity process on a network of size $y$
results in a remaining giant component consisting of a fraction $g_p(y)$ which is a fraction
$\phi=y g_p(y)$ of the $N$ nodes in the original network.

The state of the network at steady state is given, according to \cite{Bashan}, by the two equations:
$y=pg_D(x)$ and $x=pg_p(y)$, which can be reduced to a single equation:

\begin{equation}
\label{equation_x}
x=pg_p(pg_D(x)).
\end{equation}

\noindent Solving equation (\ref{equation_x}) we obtain the size of the
network at the end of a cascade initiated after random removal of $1-p$ of the nodes.

This formalism is general for any network having connectivity and dependency links.
In Sect. \ref{sec:ConnProc} we evaluate explicitly the function $g_p$, describing the connectivity
process and in Sect. \ref{sec:DepProc} we evaluate the function $g_D$, describing the action of the
dependency links.

\section{Connectivity process}
\label{sec:ConnProc}
The percolation process
can be solved analytically using the apparatus
of generating functions. We introduce the generating
function of the degree distribution $G_{0}(\xi)=\sum_k P(k) \xi^k$, where $P(k)$ is the
probability of a node to have $k$ connectivity links \cite{Newman,Shao,Shao09}.
Analogously, we introduce the generating function of the underlining
branching processes, $G_{1}(\xi)=G'_{0}(\xi)/ G'_{0}(1)$.
Random removal of a fraction $1-T$ of nodes will change the degree
distribution of the remaining nodes, so the generating functions of the
new distribution are equal to the generating functions of the original
distribution with the argument equal to $1-T(1-\xi)$
\cite{Newman}. Thus, the fraction of nodes that belong to the giant
component after the removal of $1-T$ nodes is \cite{Shao,Shao09}:
\begin{equation}
g_p(T)=1-G_{0}[1-T(1-u)],
\label{e:g}
\end{equation}
\noindent where $u=u(T)$ satisfies the self-consistency relation
\begin{equation}
u=G_{1}[1-T(1-u)].
\label{e:u}
\end{equation}

Eqs. (\ref{e:g}) and (\ref{e:u}) describe generally random networks having any degree distribution.
Specifically, we analyze two random network models that can be solved explicitly. In \ref{sec:ER} we consider
ER networks, whose connectivity degrees are Poisson-distributed \cite{er1,er1a,bollo}, and in \ref{sec:RR} we consider
RR networks, where all nodes have the same connectivity degrees $k$.

\subsection{Erd$\ddot{o}$s-R$\acute{e}$nyi (ER) networks}
\label{sec:ER}
Erd$\ddot{o}$s-R$\acute{e}$nyi (ER) networks are networks with a Poissonian degree distribution,
\begin{equation}\label{ER_dist}
    P(k')=\frac{k^{k'}}{{k'}!}e^{-k},
\end{equation}
where $k'$ is the degree of the node and $k$ is the mean degree.
For ER network with average degree $k$ the generating functions are \cite{Newman2001},
\begin{equation}\label{ER_G0_G1}
    G_{0}(\xi)=G_{1}(\xi)=\exp[k(\xi-1)].
\end{equation}
Thus, according to Eqs. (\ref{e:g}) and (\ref{e:u}),
\begin{equation}\label{e:g_p}
    g_p(T) = 1-u
\end{equation}
where $u$ satisfies the self consistent equation
\begin{equation}\label{e:u_2}
    u = e^{-kT(1-u)}.
\end{equation}

The steady state at the end of the cascade, given in Eq. (\ref{equation_x}),
becomes
\begin{eqnarray} \label{e:ER_ss}
  x &=& p(1-u) \\
  \nonumber u &=& e^{-kpg_D(x)(1-u)},
\end{eqnarray}
which can be reduced to a single equation for $u$
\begin{equation}\label{e:trans_u_1}
    u=e^{-kpg_D(p(1-u))(1-u)}.
\end{equation}

In order to present $u$, obtained from Eq. (\ref{e:trans_u_1}), in terms of $P_\infty$ recall that
 the size of the giant component of the network at the steady state, $\phi_\infty$, is given by
$\phi_\infty=xg_D(x)$, where $x$ is the solution of Eq. (\ref{e:ER_ss}).
The remaining fraction of the network at the end of the cascade, $P_\infty$, out of the initial fraction $p$
is given by $P_\infty \equiv \phi_\infty / p$. Thus, we get the relation
  $P_\infty p=xg_D(x)$.
Finally, using (\ref{e:ER_ss}) a simple equation for $P_\infty$ is obtained
\begin{equation}\label{P_inf_ln_u}
    P_\infty=-\frac{\ln u}{kp} \quad ,
\end{equation}
where $u$ is the solution of Eq. (\ref{e:trans_u_1}).

\subsection{Random-Regular (RR) networks}
\label{sec:RR}
IN RR networks all nodes have the same connectivity degree $k$.
In this case the generating function
of the degree distribution is  $G_0(\xi)=\xi^k$ and the generating function
of the underlying branching process is
$G_1(\xi)=\xi^{k-1}$, thus, $G_0(\xi)=[G_1(\xi)]^{\frac{k}{k-1}}$.
According to Eqs. (\ref{e:g}) and (\ref{e:u})
\begin{equation}\label{e:RR_gp}
    g_p(T)=1-u^{\frac{k}{k-1}},
\end{equation}
where $u$ satisfies the self consistent equation
\begin{equation}\label{e:RR_gp_u}
    u=[1+(u-1)T]^{k-1}.
\end{equation}

The steady state at the end of the cascading process is given by Eq. (\ref{equation_x})
\begin{eqnarray}\label{e:RR_gD}
    x &=& p(1-u^{\frac{k}{k-1}}) \\
 \nonumber   u &=& [1+(u-1)pg_D(x)]^{k-1}.
\end{eqnarray}
Once system (\ref{e:RR_gD}) is solved, the size of the giant component is given by $P_\infty =xg_D(x)/p$.

\section{Dependency process}
\label{sec:DepProc}

%
In the general case, when dependency links exist, each node belongs to a dependency group of size $s$ with a
probability $q(s)$ so that the number of groups of size $s$ is equal to $q(s)N/s$.
Since after random removal of $1-T$ of the nodes each group of size $s$ remains functional
with a probability $T^s$, the total number of nodes that remain functional
is given by $\sum_{s=1}^{\infty}q(s)NT^s$. Thus, we define the function $g_D(T)$ as
the fraction of nodes that remain functional out of the $TN$ nodes that were not removed,
\begin{equation}\label{gD}
    g_D(T)\equiv\sum_{s=1}^{\infty}q(s)T^{s-1}.
\end{equation}
In particular, when all dependency clusters have the same size, $s$,
\begin{equation}\label{gD_fixed}
    g_D(T)=T^{s-1}.
\end{equation}

\section{ER and RR networks with dependency clusters of the same size}
\subsection{ER network with dependency clusters of the same size}
The percolation of an ER network with average connectivity degree $k$ and dependency
clusters of size $s$ is described by substituting Eq. (\ref{gD_fixed}) into (\ref{e:trans_u_1})
\begin{equation}\label{e:ER_u_fixed}
    u=e^{-kp^s(1-u)^s},
\end{equation}
where $1-p$ is the initial fraction of removed nodes.
The size of the giant cluster, $P_\infty$, is given, using Eqs. (\ref{e:ER_u_fixed}) and (\ref{P_inf_ln_u}), by \cite{Bashan},
\begin{equation}\label{e:ER_Pinf_fixed}
    P_\infty=p^{s-1}[1-exp(-kpP_\infty)]^s   \qquad  (ER).
\end{equation}
Eq. (\ref{e:ER_Pinf_fixed}) coincides for $s=1$ (a node depends only on itself, i.e., no dependency links) with the
known Erd$\ddot{o}$s-R$\acute{e}$nyi equation \cite{er1,er1a,bollo}, $ P_\infty=1-\exp{(-kpP_\infty)}$,
for a network without dependency relations. Moreover, for $s=2$, Eq. (\ref{e:ER_Pinf_fixed}) yields the result
obtained in \cite{Parshani_PNAS} for the case of dependency pairs. The cases of q(s) being Gaussian  or Poissonian
were also studied by Bashan et al \cite{Bashan}.

\begin{figure}[htp]
\centering
\includegraphics[scale=0.6]{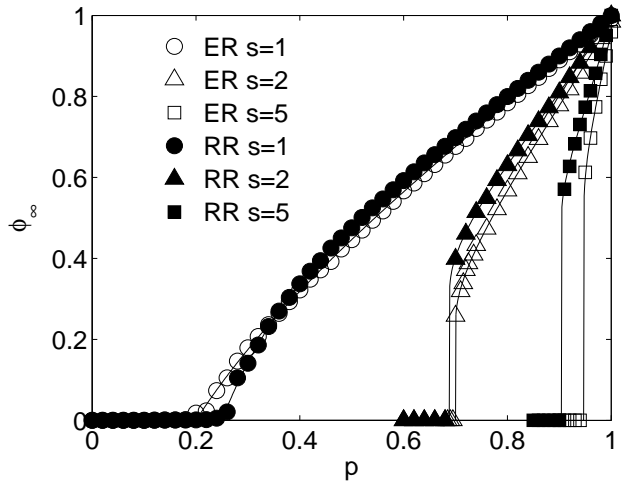}
\caption{Simulation results of the relative size of the giant component, $\phi_\infty\equiv P_\infty p$, versus $p$ for ER networks (open symbols) and
RR networks (full symbols) with connectivity degree $k=5$ compared with the theory (solid lines) as obtained from Eq.
(\ref{e:ER_Pinf_fixed}) for ER and from Eq. (\ref{e:RR_Pinf}) for RR.
For the case of
$s=1$ there are no dependency clusters and the regular percolation process
leads to the known second-order phase transition. For $s\geq2$, a first order phase transition
characterize the percolation process represented by discontinuity of $P_\infty$ at $p_c$. Both the regular second-order and the first-order percolation transitions
obey Eq. (\ref{e:ER_Pinf_fixed}) for ER and Eq. (\ref{e:RR_Pinf}) for RR.
Note that while for $s=1$ ER networks are
 more stable than RR networks ($p_c$ is smaller in ER than in RR) as $s$ increases the RR networks become
more stable compared to ER.}
\label{figure2}
\end{figure}

Fig. \ref{figure2} shows the size of the giant cluster, $\phi_\infty\equiv P_\infty p$, versus the fraction of
nodes, $p$, remaining after an initial random removal of $1-p$, for the case of ER network with fixed size of dependency clusters $s$. The case of $s=1$, where each node depends only on itself, is the known regular second order
percolation transition. For any $s\geq2$, a first order phase transition
characterizes the percolation process. Both the regular and the new first order percolation
obey Eq. (\ref{e:ER_Pinf_fixed}).

Next, we describe how to find analytically the percolation threshold, $p_c$, for the case of ER network with a fixed size $s$ of
 dependency clusters \cite{Bashan}.
Eq. (\ref{e:ER_u_fixed}), which is the condition for a steady state, have a trivial solution at $u=1$, which corresponds, by (\ref{P_inf_ln_u}),
 to a complete fragmentation of the network. For large $p$ there is another solution of $0<u<1$, corresponding to the existence
 of a giant component being a finite fraction of the network.
Therefore, the critical case corresponds to satisfying both the tangential condition for Eq. (\ref{e:ER_u_fixed}),
\begin{equation}\label{derivative_u_fixed}
    1= kp^{s}su(1-u)^{s-1} ,
\end{equation}
as well as Eq. (\ref{e:ER_u_fixed}). Thus, combining Eqs. (\ref{derivative_u_fixed})
and (\ref{e:ER_u_fixed}) we obtain a closed-form expression for the critical value, $u_c$,

\begin{equation}\label{critical_u_fixed}
        u_{c}=\exp{\left(\frac{u_{c}-1}{su_{c}}\right)}.
\end{equation}
Once $u_c$ is found, we obtain $p_c$ by substituting it into Eq. (\ref{derivative_u_fixed}),

\begin{equation}\label{p_c}
    p_{c}=\left[ksu_{c}(1-u_{c})^{s-1} \right]^{-1/s}.
\end{equation}

For $s=1$ we obtain the known result $p_c=1/k$ of Erd$\ddot{o}$s-R$\acute{e}$nyi
\cite{er1,er1a,bollo}.
 Substituting
$s=2$ in Eqs. (\ref{critical_u_fixed}) and (\ref{p_c}) one obtains ${p_c}^2=1/[2ku_c(1-u_c)]$,
which coincides with the exact result found in ~\cite{Parshani_PNAS}.

\begin{figure}[ht]
\centering
\subfigure[ER networks.]{
\includegraphics[scale=0.35]{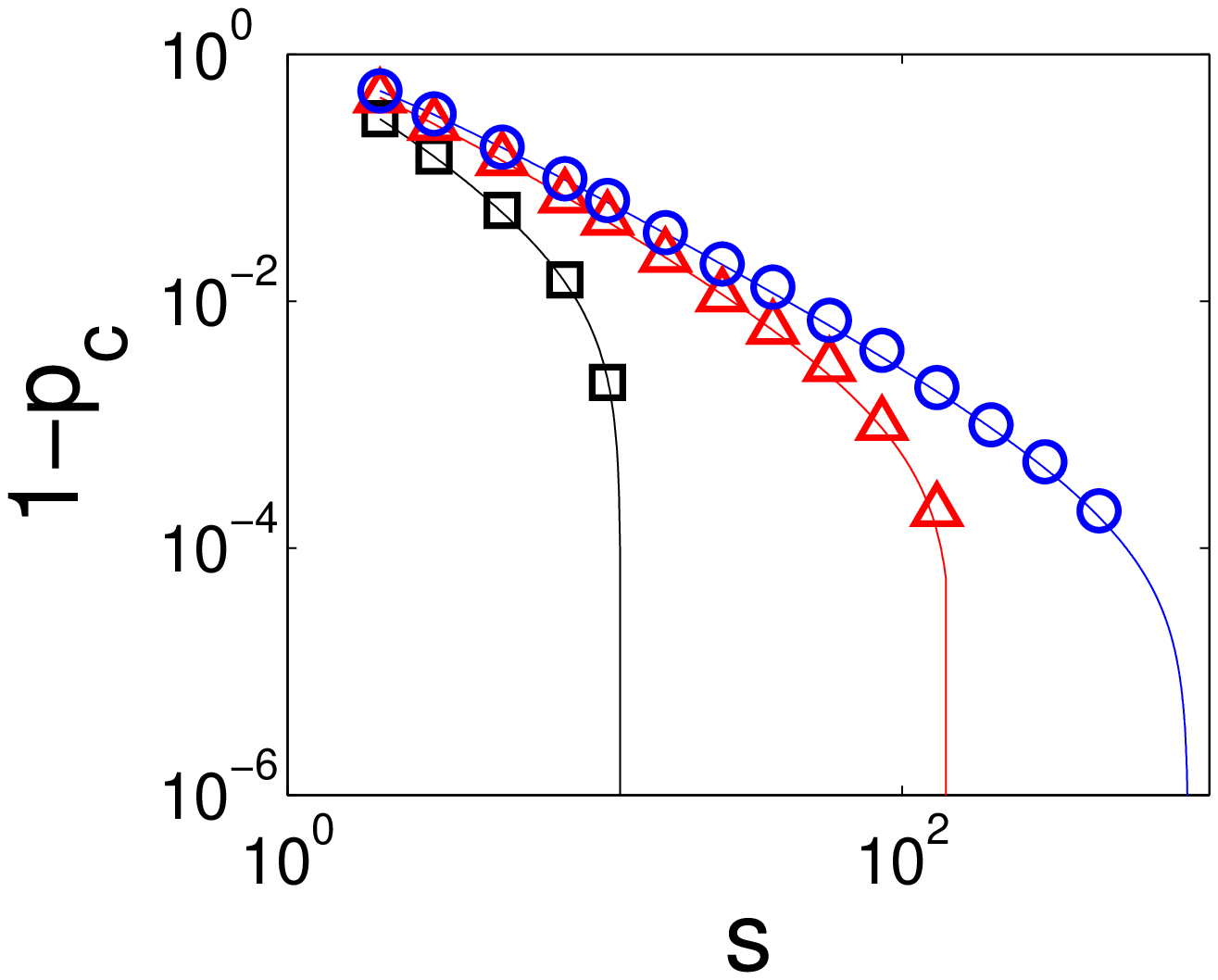}
\label{figure3_ER}
}
\subfigure[RR networks.]{
\includegraphics[scale=0.35]{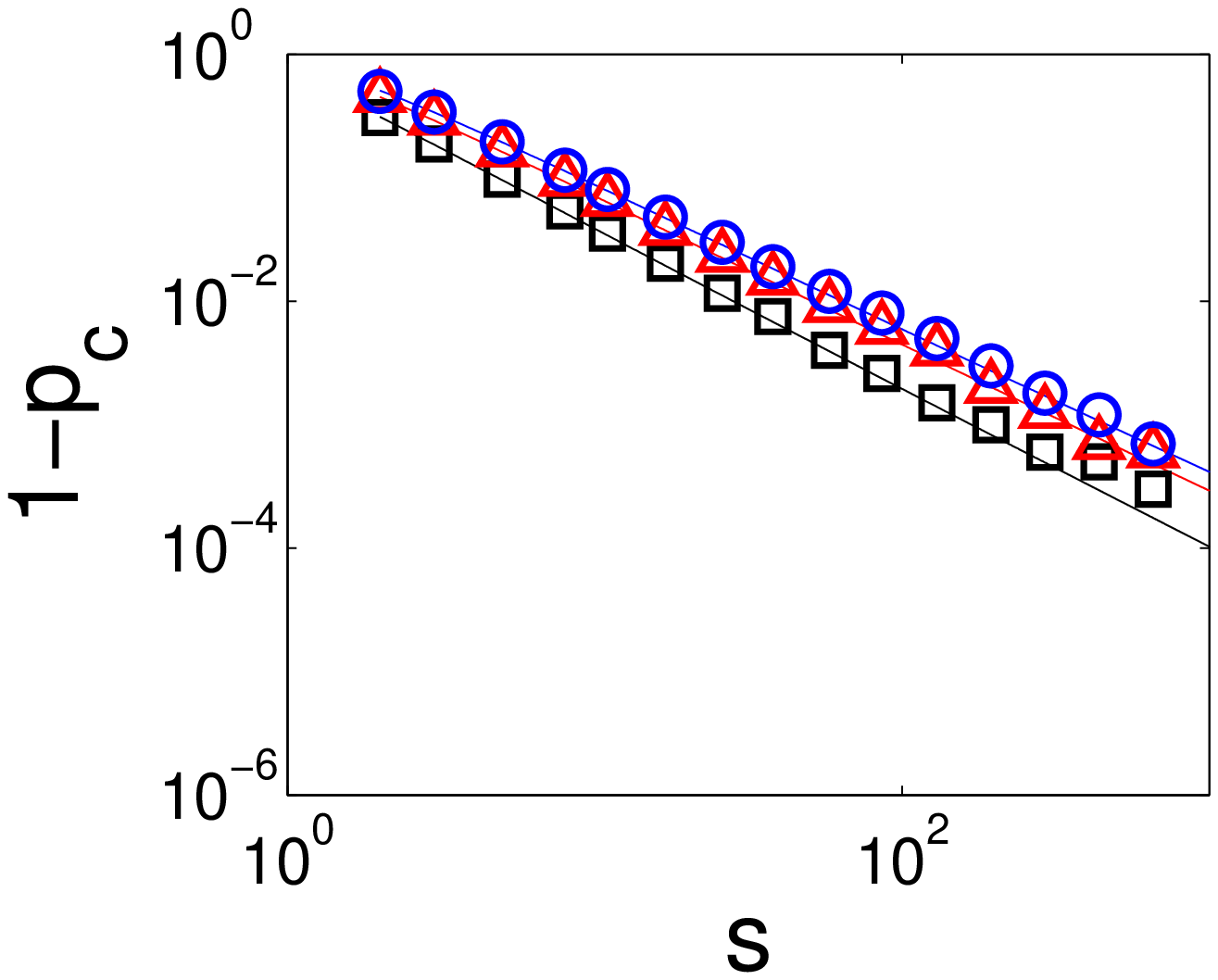}
\label{figure3_RR}
}
\caption{(Color online) Critical probability $p_c$ for (a) ER and (b) RR networks with dependency
clusters of size $s$, for different connectivity degree $k$: (black) squares for $k=5$,
 (red) triangles for $k=8$ and (blue) circles for $k=10$.
While for ER [Fig. \ref{figure3_ER}] $1-p_c$ drastically drops to zero ($p_c=1$) at $s_{max}$,
in RR [Fig. \ref{figure3_RR}] $1-p_c$ scale as power law, see Eq. (\ref{e:RR_pc_large_s}), and never reaches zero.}
\label{Pcvss}
\end{figure}

When $p_c\geq1$, removal of
even a finite number (zero fraction) of the network nodes will trigger
 a cascade of failures that fragments the whole network.
Moreover, for any given $s$ there is a minimal degree value,
\begin{equation}\label{e:ER_k_min}
    k_{min}(s)=\left[su_c(1-u_c)^{s-1} \right]^{-1},
 \end{equation}
such that $p_c$ for an ER network
with $k\leq k_{min}$ is equal to one and the network is extremely unstable.
Similarly, for any given connectivity degree $k$ there exists $s_{max}$, given by Eq. (\ref{p_c})
under the condition of $p_c=1$, such that for $s>s_{max}$, $p_c>1$ (see Fig. \ref{figure3_ER}).
 In Fig. \ref{Pcvss} we plot the values of $p_c$ as a function of $s$ for several $k$ values.

\subsection{RR network with dependency clusters of the same size $s$ }
In the case of RR network with fixed size $s$ of dependency clusters, Eqs. (\ref{e:RR_gD})
and (\ref{gD_fixed}) can be reduced to
\begin{equation}\label{e:RR_u_fixed}
    u=[(u-1)p^s(1-u^{\frac{k}{k-1}})^{s-1}+1]^{k-1}.
\end{equation}
Introducing a new variable $r\equiv u^{\frac{1}{k-1}}$, Eq. (\ref{e:RR_u_fixed}) becomes
\begin{equation}\label{e:RR_r_fixed}
    r=p^s(r^{k-1}-1)(1-r^k)^{s-1}+1.
\end{equation}
The size of the giant cluster, $P_\infty$, after random removal of $1-p$ of the network nodes,
is given by
\begin{equation}\label{e:RR_Pinf}
    P_\infty=p^{s-1}(1-r^k)^s \qquad (RR). 
\end{equation}

Similar to the case of ER networks, as shown in Fig. \ref{figure2}, the percolation process of RR networks,
given in Eqs. (\ref{e:RR_r_fixed},\ref{e:RR_Pinf}), represents a phase transition. For $p>p_c$ the size of the
network at the end of the cascade process, $P_\infty$, is finite, while for $p<p_c$ the network becomes
completely fragmented. The transition point is given by the tangential condition of Eq. (\ref{e:RR_r_fixed})
\begin{equation}\label{e:RR_rc}
    1=(1-r_c)r_c^{k-2}\left[\frac{(s-1)kr_c}{1-r_c^k}+\frac{k-1}{1-r_c^{k-1}}\right].
\end{equation}

Solving Eq. (\ref{e:RR_rc}) we obtain the transition point, $p_c$,
\begin{equation}\label{e:RR_pc}
    p_c^s = \frac{1-r_c}{(1-r_c^{k-1})(1-r_c^k)^{s-1}}.
\end{equation}
The size of the giant cluster, $P_\infty$, at $p_c$ is obtained by substituting $r_c$ into
 Eq. (\ref{e:RR_Pinf}).

Next we show analytically that for RR networks $p_c$ is always smaller than $1$.
This is in contrast to ER networks where $p_c$ can be equal or larger
 than $1$, (see Eq. (\ref{e:ER_k_min})). As shown in Fig. (\ref{Pcvss}), as the size of the dependency clusters, $s$, increases $p_c$ increases and
the network becomes more vulnerable. When the connectivity degree, $k$, increases
the system becomes more robust and $p_c$ becomes smaller.
In order to check if RR networks can become totally fragmented in removal of zero fraction of its nodes,
namely,  $p_c\geq1$,
we evaluate $p_c$ in the limit of large $s$. For $s\gg1$, $r_c$ is evaluated from Eq. (\ref{e:RR_rc})
\begin{equation}\label{e:RR_rc_large_s}
    r_c|_{s\gg1}=\left(\frac{1}{sk}\right)^{\frac{1}{k-1}}.
\end{equation}
Substituting $r_c$ into Eq. (\ref{e:RR_pc}) and taking $s\gg1$ we obtain
\begin{equation}\label{e:RR_pc_large_s}
    p_c|_{s\gg1} \simeq \left[1-\left(1-\frac{1}{k}\right)\left(\frac{1}{sk}\right)^{\frac{1}{k-1}}\right]^{\frac{1}{s}}
    \simeq 1-\left(1-1/k\right)k^{\frac{-1}{k-1}}s^{\frac{-k}{k-1}}.
\end{equation}
This formula can explain the power law behavior as a function of $s$ and the slopes seen in Fig. \ref{figure3_RR}.
Since $0<r_c<1$ and $k>1$, we obtained that the critical point of the percolation transition
 $p_c$ of RR networks is always smaller than $1$ for any given $s$.

\section{Summary}
The synergy between the connectivity and the dependency processes in
 networks having links of two types, connectivity and dependency links,
leads to cascade of failures that change the percolation properties of the
system. The new percolation laws, given by Eqs. (\ref{e:ER_Pinf_fixed}) and (\ref{e:RR_Pinf})
 for ER and RR networks respectively, predict for both ER and RR networks
a second-order percolation transition where no dependency clusters are present
 and a first-order transition for networks with dependency clusters.
However, the two topologies are dramatically different regarding their stability
in the case of large dependency clusters.
In ER networks with low connectivity degree or large dependency clusters, removal of
even a finite number (zero fraction) of the network nodes may trigger
 a cascade of failures that fragments the whole network.
  This is in contrast to RR networks where
 such cascades can be triggered only by removal of a finite fraction of nodes in the network.


%
%

\begin{acknowledgements}
We thank Yanqing Hu for helpful discussions. We thank the European EPIWORK project,
 the Israel Science Foundation, ONR, DFG, and DTRA for financial support.
\end{acknowledgements}

\end{document}